\input phyzzx.tex
\tolerance=1000
\voffset=-0.0cm
\hoffset=0.7cm
\sequentialequations
\def\rl{\rightline}

\def\t1{{\tilde 1}}

\def\t{\theta}

\REF{\MAP}{C. L. Bennett et. al., astro-ph/0302207; G. Hinshaw et. al., astro-ph/0302217; A. Kogut et. al., astro-ph/0302213.}
\REF{\GUT}{A. H. Guth, Phys. Rev {\bf D23} (1981) 347.}
\REF{\LIN}{A. D. Linde, Phys. Lett. {\bf B108} (1982) 389.}
\REF{\ALB}{A. Albrecht and P. J. Steinhardt, Phys. Rev. Lett. {\bf 48} (1982) 1220.}
\REF{\DSS}{G. Dvali, Q. Shafi and S. Solganik, hep-th/0105203.}
\REF{\BUR}{C. P. Burgess at. al. JHEP {\bf 07} (2001) 047, hep-th/0105204.}
\REF{\ALE}{S. H. Alexander, Phys. Rev {\bf D65} (2002) 023507, hep-th/0105032.}
\REF{\EDI}{E. Halyo, hep-ph/0105341.}
\REF{\SHI}{G. Shiu and S.-H. H. Tye, Phys. Lett. {\bf B516} (2001) 421, hep-th/0106274.}
\REF{\CAR}{C. Herdeiro, S. Hirano and R. Kallosh, JHEP {\bf 0112} (2001) 027, hep-th/0110271.}
\REF{\SHA}{B. S. Kyae and Q. Shafi, Phys. Lett. {\bf B526} (2002) 379, hep-ph/0111101.}
\REF{\BEL}{J. Garcia-Bellido, R. Rabadan and F. Zamora, JHEP {\bf 01} (2002) 036, hep-th/0112147.}
\REF{\KAL}{K. Dasgupta, C. Herdeiro, S. Hirano and R. Kallosh, Phys. Rev. {\bf D65} (2002) 126002, hep-th/0203019.}
\REF{\TYE}{N. Jones, H. Stoica and S. H. Tye, JHEP {\bf 0207} (2002) 051, hep-th/0203163.}
\REF{\OTH}{T. Matsua, hep-th/0302035; hep-th/0302078; T. Sato, hep-th/0304237; Y. Piao, X. Zhang and Y. Zhang, hep-th/0305171; C. P. Burgess, 
J. Cline and R. Holman, hep-th/0306079.}
\REF{\CHA}{A. D. Linde, Phys. Lett. {\bf B129} (1983) 177; Phys. Lett. {\bf B175} (1986) 395.}
\REF{\HYB}{A. D. Linde, Phys. Lett. {\bf B259} (1991) 38; Phys. Rev. {\bf D49} (1994) 748.}
\REF{\DTE}{E. Halyo, Phys. Lett. {\bf B387} (1996) 43, hep-ph/9606423.} 
\REF{\BIN}{P. Binetruy and G. Dvali, Phys. Lett. {\bf B450} (1996) 241, hep-ph/9606342.}
\REF{\TYP}{E. Halyo, Phys. Lett. {\bf B454} (1999) 223, hep-ph/9901302.}
\REF{\BRA}{E. Halyo, Phys. Lett. {\bf B461} (1999) 109, hep-ph/9905244; JHEP {\bf 9909} (1999) 012, hep-ph/9907223.}
\REF{\HAN}{A. Hanany and E. Witten, Nucl. Phys. {\bf B492} (1997) 152, hep-th/9611230.}
\REF{\BAR}{J. L. F. Barbon, Phys. Lett. {\bf B402} (1997) 59, hep-th/9703051.}
\REF{\KUT}{A. Giveon and D. Kutasov, Rev. Mod. Phys. {\bf 71} (1999) 983, hep-th/9802067.}
\REF{\BRO}{J. H. Brodie, hep-th/0101115.}
\REF{\DAS}{K. Dasgupta and S. Mukhi, Nucl. Phys. {\bf B551} (1999) 204, hep-th/9811139.}
\REF{\REN}{R. Kallosh, hep-th/0109168.}
\REF{\QUI}{B. Ratra and P. J. E. Peebles, Phys. Rev. {\bf D37} (1988) 3406.}
\REF{\STE}{L. Wang, R. R. Caldwell, J. P. Ostriker and P. J. Steinhardt, Astrophys. {\bf J530} (2000) 17, astro-ph/9901388.}
\REF{\TEV}{E. Halyo, JHEP {\bf 0110} (2001) 025, hep-th/0105216.}
\REF{\STR}{A. Strominger, JHEP {\bf 0110} (22001) 034; hep-th/0106113.}
\REF{\INF}{A. Strominger, hep-th/0110087.}
\REF{\HOL}{E. Halyo, hep-th/0203235.}

\singlespace
\rl{hep-th/0307223}
\rl{\today}
\pagenumber=0
\normalspace
\medskip
\bigskip
\titlestyle{\bf{Models of Inflation on D--Branes}}
\smallskip
\author{ Edi Halyo{\footnote*{e--mail address: vhalyo@stanford.edu}}}
\smallskip
 \centerline{California Institute for Physics and Astrophysics}
\centerline{366 Cambridge St.}
\centerline{Palo Alto, CA 94306}
\smallskip
\vskip 2 cm
\titlestyle{\bf ABSTRACT}
We obtain models of chaotic, slow--roll, hybrid and D--term inflation from the Hanany--Witten brane configuration and its deformations. 
The deformations are given by the different orientations of the branes and control the parametes of the scalar potential such as the inflaton mass, 
Yukawa couplings and the anomalous D--term. The different inflationary models are continuously connected to each other and arise in different limits 
of the parameter space. We describe a compactified version of the brane construction that also leads to models of inflation. 

\singlespace
\vskip 0.5cm
\endpage
\normalspace

\centerline{\bf 1. Introduction}
\medskip

The recent WMAP data[\MAP] seems to vindicate the inflationary paradigm[\GUT-\ALB] according to which the universe went through an exponential expansion during its very early stages. 
Since such 
early times correspond to very high energies, models of inflation need to be realized in theories which describe physics at such energies. The most fundamental and promising of these
theories is string theory which is a theory of quantum gravity. Therefore, it is important to be able to build models of inflation in string theory. 

Considerable progress has been made in this direction by realizing inflation on D--branes which are nonperturbative states of string theory[\DSS-\OTH]. These D--brane inflation models 
come in two types: those realized either by $Dp$-$\bar Dp$ branes or by
$Dp$-$D(p+2)$ branes. The inflaton corresponds either to the distance or to the angle between the respective branes. The slow--roll of the inflaton during the inflationary era corresponds
to the slow motion of the branes towards each other. Inflation ends when the branes are very close to each other so that a tachyon is formed which later condenses.

In this paper we consider the Hanany--Witten brane construction[\HAN] and its deformations in order to obtain four different models of D--brane inflation, namely chaotic, slow--roll, 
hybrid and D--term inflation. The Hanany--Witten construction is given by a $D4$ brane along directions $1236$ stretched between two $NS5$ branes along the $12345$ directions. 
In addition, there 
is a $D6$ brane along directions $123789$. Our world lies along directions $123$ which are common to all the branes.
The deformations of the Hanany--Witten model in question are rotations of the $D4$ and $NS5$ branes used in the construction. A relative angle ($\theta_1$) between the 
$NS5$ branes gives an inflaton mass term. An overall angle (different than $\pi/2$) between both $NS5$ branes and the $D6$ brane ($\theta_2$) gives nonzero Yukawa couplings, 
i.e. the couplings between the different scalars of the model. 
In addition, one can consider the rotation of the $D4$ brane relative to the $D6$ brane ($\theta_3$) which leads to an anomalous D--term. We show that brane configuration with 
$\theta_1 \not= 0$,
$\theta_2=\pi/2$ and $\theta_3=0$ leads to chaotic inflation. A configuration with $\theta_1=0$, $\theta_2=\pi/2$ and $\theta_3 \not= 0$ gives slow--roll inflation. 
When, $\theta_1$ and $\theta_3$ are nonzero and $\theta_2 \not= \pi/2$ we get hybrid inflation. Finally, if $\theta_1=0$, $\theta_2 \not= \pi/2$ and $\theta_3 \not= 0$ we get D-term inflation.

The picture that emerges is a very appealing one. Chaotic, slow--roll, hybrid and D--term inflation models on D--branes can be obtained from four brane configurations which are 
related to each other by continuous deformations (rotations of the branes). Each inflation model is obtained in a different limit of the parameter or configuration space of the D--brane 
construction and is continuously 
connected to the others. As mentioned above, these different inflation models are obtained when one, (a different) one, two and three angles are nonzero. 
In this sense, the Hanany--Witten model and its deformations provide a unified microscopic framework for the different models of inflation.
Usually, different models of inflation 
require different (inflaton) potentials and therefore are not related in any sense. The initial conditions determine whether inflation takes place or not but not
the type of inflation possible since this is fixed by the potential which describes the underlying theory. In our brane construction, the deformations control the value of the 
parameters that appear in
the scalar potential. As a result, we can obtain many different scalar potentials from the same basic brane construction which can lead to different inflation models.
In this case, it seems that the type of inflation that occurs is also a question of initial conditions which fix the deformation parameters.

The main shortcoming of the above construction is the difficulty in compactifying the extra six dimensions which are parallel to some of the branes. In fact a compactified version
of the Hanany--Witten construction does not exist. However, there is another brane 
construction with six compact dimensions which 
is (loosely) related to the above construction by T-duality. In this construction, one considers type IIB string theory compactified on an orientifold of $K3 \times T^2$. This manifold
is very similar to the T dual of the two NS5 branes at a distance, namely $K3 \times T^2$. After T duality, the $D4$ and $D6$ branes become $D3$ and $D7$ branes. 
In order to cancel the overall $D7$ branes charge one needs to include orientifolds, i.e. $O7$ planes. What happens to the three deformations of the original brane construction, namely
the three types of rotations considered above in this new framework? Standard T duality shows that a relative angle between the $D4$ and $D6$ branes becomes a flux on the world--volume
of the $D7$ brane. The relative angle between the two $NS5$ branes becomes a geometrical deformation of the $K3 \times T^2$ towards a conifold (which means a deformation of the $T^2$
to a sphere, $P^1$). The overall rotation of both $NS5$ branes interchanges the planes $45$ and $89$ and therefore mixes two directions along the $K3$ with those of the $T^2$. 
In this construction, inflation arises due to the slow motion or rotation of the $D3$ brane relative to the $D7$ brane. The two NS5 branes become the background geometry after the T duality.
We can argue that we obtain the four models of inflation from T duality but it is quite difficult to follow the scalar potential under these complicated geometrical deformations.
In this alternative construction, the price we pay for compactification is that the original simple deformations (rotations) become fluxes and complicated geometric deformations which 
are more difficult to visualize and quantify.

This paper is organized as follows. In the next section we briefly review the models of chaotic, slow--roll, hybrid and D--term inflation and the conditions for acceptable inflation
according to the WMAP data. In section 3, we describe the Hanany--Witten 
brane construction we use and its deformations by the relative rotations of the branes. In section 4, we describe how the different models of inflation arise from the different 
deformations of the Hanany--Witten brane construction. 
In section 5, we consider a compactifed D--brane construction which gives the different models of inflation and is loosely related to the one in section 3 by T duality.
Section 6 contains our conclusions and a discussion of our results.

\bigskip
\centerline{\bf 2. Models of Inflation}
\medskip

In this section we briefly review some well--known models of inflation, namely chaotic, slow--roll, hybrid and D-term inflationary models. As we will see in section 4, 
these are the models of inflation that can be easily obtained from the different deformations of the Hanany-Witten brane model.

All models of inflation have to satisfy the recently obtained WMAP constraints[\MAP]. These include the slow--roll constraints $0<\epsilon_1<0.022$ and 
$-0.06<\epsilon_2<0.05$ where $\epsilon_1=\epsilon$
and $\epsilon_2=2(\epsilon-\eta)$ which are defined by the slow--roll parameters
$$\epsilon={M_P^2 \over 2} \left(V^{\prime} \over V \right)^2 \eqno(1)$$
and
$$\eta=M_P^2 \left(V^{\prime \prime} \over V \right) \eqno(2)$$
Inflation occurs as long as these conditions are satisfied and ends when at least one of them is violated.
The models also have to produce the correct amount of scalar density perturbations; $18.8\times 10^{-10}<A_s<24.8 \times 10^{-10}$ where
$$A_s={H^2 \over {8 \pi^2 M_P^2 \epsilon_1}} \eqno(3)$$
$A_s$ is related to the density perturbations by $\delta \rho/\rho \sim A_s^2$.
The ratio of the amplitudes for the tensor and scalar perturbations must satisfy $0<R<0.35$ where $R \sim 16 \epsilon_1$. The scalar and tensor spectral indices which parametrize
the deviations from scale invariant perturbations are constrained by
$0.94<n_s<1.02$ and $-0.044<n_t<0$ where
$$n_s \sim 1-2 \epsilon_1 -\epsilon_2 \qquad \qquad n_t \sim -2\epsilon_2 \eqno(4)$$
In addition, the models must result in about 60 e--folds of inflation
$$N = M_P^{-2} \int {V \over V^{\prime}} d\phi \sim 60 \eqno(5)$$
These conditions impose stringent limits on the parameters of inflationary models.

Chaotic inflation: Chaotic inflation[\CHA] is the simplest of all inflationary models with a very simple one field potential
$$V(\phi)= {1 \over 2}m^2 \phi^2 \eqno(6)$$
or any other power law potential for the inflaton. 
The slow--roll constraints imply a very large inflaton value $\phi> 7M_P$ during inflation. For such large inflaton values, the inflaton rolls down its potential slowly
due to the large Hubble constant (compared to the inflaton mass). Inflation ends when the inflaton rolls down to $\phi \sim M_P$. Afterwards the inflaton rolls very fast to the 
minimum of its potential at $\phi=0$.
The requirement for 60 e--folds gives $\phi \sim 10M_P$.
The proper amount of scalar density perturbations implies a very small inflaton mass, 
$m \sim 5 \times 10^{-5}M_P$. Both of these conditions are usually problematic in field theory since the inflaton mass requires fine tuning and its value is such that 
field theory is not applicable. We will see that both of these problems are naturally resolved in chaotic inflation that arises on branes.

Slow--roll inflation: Slow--roll inflation[\LIN-\ALB] is probably the most generic model in which the inflaton can have any potential that satisfies the slow--roll contraints 
mentioned above. In this case, the slope of the inflaton potential is very small during the inflationary era
and therefore the inflaton is almost constant. The simplest inflaton potential that one can have is
$$V(\phi)=g^2(v^2-\phi^2)^2 \eqno(7)$$
which is the generic scalar potential that leads to spontaneous symmetry breaking. At a certain value of the inflaton, the slow--roll conditions are violated and inflation ends.
Afterwards the inflaton rolls very fast to the minimum of its potential at $\phi=v$. The slow-roll conditions impose constraints on the parameters of the potential,
$g$ and $v$.

Hybrid inflation: Hybrid inflation[\HYB] requires two fields, the inflaton ($\phi$) and the trigger ($\psi$) fields with a potential
$$V(\phi, \psi)=g^2(v^2-\psi^2)^2+m^2\phi^2+\lambda^2 \phi^2\psi^2 \eqno(8)$$
For large initial values of the field $\phi>\sqrt 2 gv$ the minimum of the potential for the trigger field is at $\psi=0$. When $gv>H$, $\psi$ rolls down very quickly and 
settles down to its minimum.
On the other hand, for a small inflaton mass and small values of $\psi$, the inflaton will roll down slowly from its large initial value to its minimum at $\phi=0$. This slow--roll 
corresponds to inflation. Once $\phi<\sqrt 2 gv$, $\psi=0$ becomes a local maximum and a new minimum appears at $\psi=v$. The trigger field rolls down to the new minimum.
While the trigger field is rolling down to $\psi=v$ the inflaton mass increases and the Hubble constant decreases so that at some point inflation ends. Finally $\psi$ settles down
to $\psi=v$ (and $\phi$ to $\phi=0$) and the vacuum energy vanishes.
The WMAP constraints in eqs. (1)-(5) impose conditions on the parameters of the model, $g,v,m$ and the initial values for $\phi$ and $\psi$.

D--term inflation: D--term inflation[\DTE-\BRA] is a variant of hybrid inflation in which the trigger field potential arises from an anomalous D--term and there is no tree level inflaton mass.
The tree level potential is
$$V(\phi, \psi)=g^2(v^2-\psi^2)^2+\lambda^2 \phi^2\psi^2 \eqno(9)$$
D--term inflation proceeds exactly as hybrid inflation with the difference that the inflaton potential is flat at tree level.
The inflaton gets a small mass from one--loop effects due to supersymmetry breaking during the inflationary era
$$V_{1-loop}=g^2v^4 \left(1+{g^2 \over {16 \pi^2}} log{\lambda \phi \over M_P} \right) \eqno(10)$$
and rolls down slowly to its minimum at $\phi=0$. The rest of the scenario proceeds as in hybrid inflation.

Usually, these inflation models arise in different models with different scalar potentials. In fact the models in question are built in order to realize the specific inflaton
potential. What we show below is that these four models of inflation can arise from the same microscopic model which is the Hanany--Witten brane construction and its deformations.
Thus, the brane construction below can be viewed as unifying microscopic framework for inflationary models.

\bigskip
\centerline{\bf 3. D--Brane Constructions}
\medskip

The basic brane constructions that we use in order to realize the different inflation models are the well--known Hanany--Witten model and its deformations[\HAN, \KUT]. 
In the Hanany--Witten construction there is a $D4$ brane along directions $1236$ stretched between two $NS5$ branes along $12345$ directions. In addition, there is a $D6$ brane 
along directions $123789$. Our world
lies along directions $123$ which are common to all the branes. The other six directions need to be compactified in order to have four dimensional gravity which is very problematic 
in the context of the Hanany--Witten construction. We will discuss a compactified version of a dual model in section 5. 

For energies lower than the inverse of the 
distance between the two $NS5$ branes along the $6$ direction, $L$, the field theory on the $D4$ branes is a $1+3$ dimensional, ${\cal N}=2$ supersymmetric $U(1)$ gauge theory with a 
charged hypermultiplet. In terms of ${\cal N}=1$ supersymmetry, these are a gauge multiplet with one neutral and two conjugate, charged chiral multiplets. 
The scalar fields of the model are given by a $U(1)$ gauge boson and three scalars: $\phi_{1,2}$ (which are conjugates with opposite charges)  
from the hypermultiplet and $\Phi$ from the vector multiplet. 
$\Phi$ arises from 4-4 strings and parametrizes the position of the $D4$ brane (relative to the $D6$ brane) along the $4$ and $5$ directions, $\Phi=(X_4+iX_5)/2\pi \ell_s^2$.
$\phi_{1,2}$ arise from 4-6 strings along the $7,8,9$ and (the Wilson loop value along the compact gauge) $6$ directions, $\phi_1=(X_8+iX_9)/2\pi \ell_s^2$ and 
$\phi_2=(X_6+iX_7)/2\pi \ell_s^2$.
${\cal N}=2$ supersymmetry allows only the superpotential
$$W=g \Phi \phi_1 \phi_2 \eqno(11)$$
Assuming that the above construction is compactified along the $456789$ directions on a $T^6$ with radius $R$ in a consistent way (so that $M_P^2=g_s^2 \ell_s^8/LR^5$), 
the gauge coupling is given by 
$$g=2\pi\sqrt{{g_s \ell_s^6} \over {LR^5}} \eqno(12)$$
The equality between the Yukawa and gauge couplings is dictated by ${\cal N}=2$ supersymmetry.
In terms of the scalars, the potential is
$$V_1=g^2{(|\phi_1|^2+|\phi_2|^2)|\Phi|^2+g^2 |\phi_1|^2|\phi_2|^2}+g^2(|\phi_1|^2-|\phi_2|^2)^2 \eqno(13)$$
where the first and second terms are the F and D terms respectively. 

There are three deformations of the above configuration that control the parameters in the scalar potential. As we will see below, these deformations are rotations of one or 
more of the branes in the Hanany--Witten construction and control the singlet scalar mass, the Yukawa coupling and the anomalous D--term.

Rotation of one $NS5$ brane:
First, one can rotate one of the $NS5$ branes keeping everything else the same[\KUT, \BAR]. Let us define the complex coordinates $v=X_4+iX_5$, $s=X_6+iX_7$ and $w=X_8+iX_9$.
In the above configuration, the $NS5$ branes are stretched along $123$ and $v$ directions whereas the $D6$ brane is along the $1237$ and $w$ directions. We can rotate one of 
the $NS5$ branes (by angle $\theta_1$) so that
$$v_{\theta}=vcos\theta_1+wsin\theta_1 \qquad \qquad  w_{\theta}=-vsin\theta_1+wcos\theta_1 \eqno(14)$$
Now the rotated $NS5$ brane (denoted by $NS5_{\theta}$) is stretched along $123$ and $v_{\theta}$ directions or at $w_{\theta}=0$ (in addition to the $X_7$ coordinate).
In the brane field theory this rotation  introduces a mass for $\Phi$ so the scalar potential gets an additional term $V_2=m_{\Phi}^2 \Phi^2/2$ where
$$m_{\Phi}={tan( \theta_1) \over {2\pi \ell_s}} \eqno(15)$$ 
A simple way to see this is to 
note that for the original configuration ($\theta_1=0$) this mass vanishes whereas for a pair of perpendicular $NS5$ branes ($\theta_1=\pi/2$) the mass diverges.
On the other hand, it is well--known that the former brane configuration gives a theory with ${\cal N}=2$ SUSY whereas the latter has only ${\cal N}=1$ SUSY. On the field theory
side the same SUSY breaking can be accomplished by giving a mass to the adjoint singlet field $\Phi$. This mass term breaks ${\cal N}=2$ SUSY down to ${\cal N}=1$ so that at 
energies lower than the $\Phi$ mass (or at any energy if this mass diverges) only ${\cal N}=1$ SUSY is visible. Thus, $m_{\Phi}$ must vanish for parallel
$NS5$ branes and diverge for perpendicular ones. Taking into account the dimension of mass we get the result in eq. (15) up to a numerical constant.
With a nonzeo mass there is a supersymmtric vacuum only at $\Phi=0$ which means the Coulomb branch has
been completely lifted. Note that even though this is a scalar mass, it can be naturally small since its value is related to an enhancement of symmetry, namely enhancement of
${\cal N}=1$ supersymmetry to ${\cal N}=2$.

Rotation of both $NS5$ branes (by a common angle):
Second, one can rotate both $NS5$ branes by the same angle $\theta_2$ as in eq. (14) above[\KUT]. In this case, we get two $NS5_{\theta_2}$ branes along $123$ and $v_{\theta_2}$ directions. 
This necessarily
rotates the $D4$ brane stretched between the $NS5$ branes. When $\theta_2=\pi/2$, the $D4$ brane ends perpendicularly on the $D6$ brane. 
The length of the (shortest) stretched 4-6 strings vanishes which in field theory corresponds to vanishing masses for the charged fields $\phi_{1,2}$. In the field 
theory, this means that a 
rotation of both $NS5$ branes by a common angle changes the Yukawa coupling in eq. (11) to $\lambda= gcos\theta_{2}$. We see that for the original configuration ($\theta_2=0$) 
the Yukawa coupling
is equal to the gauge coupling $g$ whereas for $\theta_2=\pi/2$ it vanishes. In this new configuration, $\Phi$ parametrizes the fluctuations of the $D4$ brane (or 
the fluctuations of the 4-4 strings)
along the $v_{\theta_2}$ direction and remains massless since the $NS5$ branes remain parallel. 

Rotation of the $D4$ brane relative to the $D6$ brane:
Third, one can move one of the $NS5$ branes relative to the other along some of the $7,8,9$ (and the Wilson loop $6$) directions. For example moving one $NS5$ brane relative to the other 
along the $7$ direction causes a rotation of the $D4$ brane in the $67$ plane[\KUT]. This case has been explored in detail in ref. [10]. The rotation 
(by angle $\theta_3$) introduces an anomalous Fayet--Iliopoulos D--term[\CAR]
$$\xi={L^{1/2} \over {4\pi^2 g_s^{1/2} \ell_s^{5/2}}} sin \theta_3 \eqno(16)$$
A simple way to obtain the above result is to calculate the difference in energy between the supersymmetric configuration with the $D4$ brane along $1236$ directions and the rotated
one and equate this to the energy in field theory due to the anomalous D--term[\BRO]. The original and final lengths of the $D4$ brane are $L$ and $L/cos \theta_3$. Using the $D4$ brane
tension $T_4=1/g_s \ell_s^5$ we find for small angles that $ \xi \sim L^{1/2} \theta_3/ g_s^{1/2} \ell_s^{5/2}$ which agrees with eq. (16). 
In this case, $\Phi$ remains massless and the gauge coupling is equal to the Yukawa coupling.
This rotation breaks SUSY completely. However, there is a supersymmtric vacuum with $\Phi=0$ and $\phi_2 \not=0$. In the brane picture, this state is described by the $D4$ brane
broken into two segments which are perpendicular to the $D6$ brane. 
We see that again the Coulomb branch has been completely lifted.

Taking these three deformations of the Hanany-Witten construction into account, the most general brane configuration with all three angles nonzero leads to the scalar potential
$$V=\lambda^2{(|\phi_1|^2+|\phi_2|^2)|\Phi|^2+\lambda^2 |\phi_1|^2|\phi_2|^2}+{1 \over 2}m^2\Phi^2 +g^2(|\phi_1|^2-|\phi_2|^2+{\xi \over g})^2 \eqno(17)$$

We see that the three deformations described above give us complete control over the scalar potential. Different choices of the parameters $\theta_{1,2,3}$ lead to different
potentials. In the next section, we show that we can obtain the four models of inflation described in section 2 by carefully choosing the deformation parameters.

\bigskip
\centerline{\bf 4. Models of Inflation on D--Branes}
\medskip

In this section, we show that the different types of inflation models mentioned in section 2 can be obtained from the Hanany--Witten brane configuration in different limits of the 
parameter space. Brane configurations with different orientations defined by the three angles $\theta_{1,2,3}$ introduced above lead to chaotic, slow--roll, hybrid and D--term
inflation models. Thus, these different inflationary models are continuously connected and the brane construction provides a unifying framework for them.

Chaotic inflation on D--branes: Consider the Hanany--Witten brane construction with a $D4$ brane along the $1234$ directions and stretched between two $NS5$ branes along $12389$ directions. 
We also include
a $D6$ brane along the $123789$ directions. This corresponds $\theta_1=\theta_3=0$ and $\theta_2=\pi/2$ in the notation of section 3. As a result, the superpotential in eq. (11) and 
the anomalous
D--term both vanish. In addition, the mass of the neutral singlet $\Phi$ is zero since the $NS5$ branes are parallel. If we now rotate one of the $NS5$ branes relative to the other one 
by a small angle $\alpha$ so that overall
it is rotated by $\theta_2=\pi/2-\alpha$ rather than by $\pi/2$, the singlet gets a mass $m= tan{\theta_2}/ 2 \pi \ell_s$. Thus we are left with a simple potential
$$V={1 \over 2} m^2 \Phi^2  +g^2(|\phi_1|^2-|\phi_2|^2)^2 \eqno(18)$$
which gives chaotic inflation if $\phi_1$ and $\phi_2$ are small or $g<<1$ and the parameters $m$ and $\Phi_{in}$ satisfy the WMAP constraints. 
The slow--roll conditions in eqs. (1) and (2) impose $\Phi_{in}={v/ 2\pi \ell_s^2} >10M_P$. This condition which is very problematic in field theory can be naturally satisfied in the brane
construction by taking the interbrane distance to be larger than the string length, $v> \ell_s$.
The correct amount of density perturbations
$$19 \times 10^{-10}< {1 \over {12 \pi}} {{m^2 \Phi_{in}} \over M_P^6}= {{tan^2 \theta_1 v^4} \over {64 \pi^6 \ell_s^{10} M_P^6}} <25 \times 10^{-10} \eqno(19)$$
requires a very small inflaton mass, $m \sim 10^{-5}M_P$ which looks like fine tuning. However, in this case a small mass is natural since when the mass vanishes supersymmetry
is enhanced from ${\cal N}=1$ to ${\cal N}=2$.
We also find that 
$$R < {{32 M_P^2} \over {\Phi_{in}^2}}= 128 \pi^2 {{M_P^2 \ell_s^4} \over {v^2}}<0.35 \eqno(20) $$ 
which means that the magnitude of tensor density fluctuations may be as large as $1/3$ of the scalar fluctuations.  
The spectral index is given by
$$0.94< n_s \sim 1-{4M_P^2 \over \Phi_{in}^2}= 1-16 \pi^2 {{M_P^2 \ell_s^4} \over v^2}  < 1.02 \eqno(21) $$ 

From the point of view of the brane construction, there is a net force between the $D4$ and $D6$ branes
because the two $NS5$ branes are slightly nonparallel. The $D4$ brane falls slowly towards the $D6$ brane which results in chaotic inflation. The inflaton is
the field $\Phi$ that parametrizes the distance between the $D4$ and $D6$ branes along the $v$ direction. Inflation ends when the $D4$ brane is at a distance from the $D6$ 
brane such that $\Phi=v/2 \pi \ell_s^2 \sim 10M_P$ and starts rolling fast. The bottom of the scalar potential with $\Phi=0$ which is the endpoint for inflation 
corresponds to the $D4$ brane ebbedded in the $D6$ brane. 
This is a supersymmetric configuration since the number of mutually transverse directions between the branes is four.

Slow--roll inflation on D--branes: Consider, now the deformed Hanany--Witten model with $\theta_1=0$, $\theta_2=\pi/2$ and $\theta_3\not =0$. The nonzero $\theta_3$ means that 
the $D4$ brane is rotated in the $67$ plane and
$NS5$ branes are at different positions along the $7$ direction. In this case, the singlet mass and the
superpotential vanish but there is a nonzero anomalous D--term. Thus the scalar potential is
$$V=g^2(|\phi_1|^2-|\phi_2|^2+{\xi \over g})^2 \eqno(22)$$
where $g$ and $\xi$ are defined by eqs. (12) and (16).
The above potential can give slow--roll inflation if the parameters $g$ and $v$ satisfy the constraints arising from the WMAP data which are 
$$\epsilon_1 \sim {{2 g^2 M_P^2 \phi_2^2} \over \xi^2}={{16 \pi^2 M_P^2 s^2 \ell_s^7} \over {R^5 L^2 sin^2 \theta_3}}<0.022 \eqno(23)$$
and
$$-0.06< \epsilon_2 \sim {4M_P^2 \over \xi} ({g^2 \phi_2^2 \over \xi}-1)= {{16 \pi MP^2 \ell_s^{5/2}} \over {\sqrt L sin \theta_3}} ({{4 \pi s^2 g_s^{1/2} \ell_s^{9/2}} 
\over {R^5 L^{3/2} sin \theta_3}} -1) <0.05 \eqno(24)$$
These conditions can be satisfied if the coupling $g$ is very small which can be easily obtained by taking $L>>\ell_s$ and/or $R>> \ell_s$.
In order to get enough densiy perturbations we need
$$19 \times 10^{-10} < {\xi^4 \over {6 \pi g^2 M_P^6 \phi_2^2}}= {{g_s R^5L^3} \over {128 \pi^3 \ell_s^{12} M_P^6 s^2}} < 25 \times 10^{-10} \eqno(25)$$
The ratio of the magnitudes of the tensor and scalar perturbations is
$$R \sim 32 (4 \pi)^3 g_s {{M_P^2 s^2 \ell_s^7} \over {R^5 L^2}}<0.35 \eqno(26)$$ 
The scalar spectral index is
$$0.94< n_s \sim 1-{{8 g^2 M_P^2 \phi_2^2} \over {\xi^2}}+ {4 M_P^2 \over \xi}< 1.02 \eqno(27)$$.

From the brane point of view, the $D4$ brane breaks into two segments on the $D6$ brane in order to minimize the potential energy. 
After breaking,
the $D4$ brane segments rotate slowly in a manner that makes them perpendicular to the $D6$ brane. This slow rotation corresponds to the slow--roll inflationary era.
The inflaton is the field $\phi_2$ that parametrizes the angle $\theta_3$ that the $D4$ brane makes on the $67$ plane and is related to the fluctuations of the 4--6 strings 
$\phi_2=X_6+iX_7/\ell_s^2$.
Note that $\Phi$ which parametrizes the distance between the $D6$ and $D4$ branes along the $v$ direction is absent from the potential. However, this 
does not mean that $\Phi$ is  modulus. In order be in the Higgs phase, we need $\Phi=0$ so the two branes are necessarily at the same $v$ coordinate.
Inflation ends when one of the the slow--roll conditions in eqs (23) and (24) is violated. Then the $D4$ 
brane starts to rotate fast. The bottom of the scalar potential at $\phi_2=v$
corresponds to the configuration with the $D4$ brane broken into two segments perpendicular to the $D6$ brane. This is the unique
supersymmetric brane configuration.

Hybrid inflation on D--branes: We now take the Hanany--Witten configuration and deform it by the three angles mentioned in section 3. Consider first rotating the $D4$ brane relative 
to the $D6$ brane, i.e. take $\theta_3\not =0$. This gives an anomalous D--term as in eq. (16). Second, rotate the two $NS5$ branes by a common angle $\theta_2$ which leads to a generic 
Yukawa potential with coupling $\lambda$. Finally, rotate one of the
$NS5$ branes by an additional angle $\theta_1$ in order to get a nonzero inflaton mass. The above deformations give the most general scalar potential as in eq. (17) 
$$V=\lambda^2{(|\phi_1|^2+|\phi_2|^2)|\Phi|^2+\lambda^2 |\phi_1|^2|\phi_2|^2}+{1 \over 2}m^2\Phi^2 +g^2(|\phi_1|^2-|\phi_2|^2+{\xi \over g})^2 \eqno(28)$$
This potential leads to hybrid inflation if the parameters $\lambda$, $m$, $g$ and $v$ satisfy the WMAP conditions. The slow--roll conditions are
$$\epsilon_1 \sim {{M_P^2 m^4 \Phi^2} \over { 2 \xi^4}} ={{2 M_P^2 tan^2 \theta_1} \over {\pi^2 sin^4 \theta_3}} {{v^2 g_s^2 \ell_s^2} \over {L^2}} < 0.022 \eqno(29)$$
and
$$-0.06 < \epsilon_2 \sim {{2 M_P^2 m^2} \over {\xi^2}} ({{m^2 \Phi^2} \over {2 \xi^2}}-1)={{M_P^2 \ell_s^3} \over {2g_s L}} {{tan^2 \theta_1} \over {sin^2 \theta_3}} 
({{v^2 tan^2 \theta_1} \over {8 \pi^2 sin^2 \theta_3 \ell_s L}}-1) <0.05 \eqno(30)$$
Hybrid inflation does not require any small parameters (which is one of its advatages over other models) so it is relatively easy to satisfy the above constraints.
The correct magnitude of density perturbations requires
$$19 \times 10^{-10}< {{2 \xi^6} \over {3 \pi M_P^6 m^4 \Phi^2}}={{sin^2 \theta_3} \over {1024 \pi g_s^3 tan^4 \theta_1}} {{L^3} \over {M_P^6 \ell_s^7 v^2}}< 25 \times 10^{-10} \eqno(31)$$
We also find 
$$R \sim {{16 M_P^2 m^2} \over {\xi^2}}={{64 M_P^2 g_s tan^2 \theta_1 \ell_s^3} \over {sin^2 \theta_3 L}}<0.35 \eqno(32)$$
and
$$0.94< n_s \sim1- {{2M_P^2 m^4 \Phi^2} \over {\xi^4}}+ {{2 M_P^2 m^2} \over {\xi^2}}<1.02 \eqno(33)$$
for the ratio of the magnitude of the fluctuations and the spectral index.

The initial brane configuration is the one given above with large 
distances between the branes. Initially, $\phi_2$ has a large positive mass and therefore it is stuck at the false minimum of its potential at $\phi_2=0$. $\phi_2$ is the trigger 
field that parametrizes
the distance between the $D4$ and $D6$ branes along the $7$ direction. Initially, this distance and therefore the angle $\theta_2$ between the branes remains fixed.
In the meantime the inflaton $\Phi$ that parametrizes the distance between the $D4$ and $D6$ branes rolls down slowly towards $\Phi=0$. This corresponds to the slow
motion of the $D4$ brane towards the $D6$ brane along the $v$ direction. When the distance satisfies $\Phi< \sqrt{g \xi}/\lambda$, 
$\phi_2$ becomes tachyonic and starts rolling towards $\phi_2=\sqrt{\xi/g}$. In the brane picture this corresponds to the breaking of the $D4$ brane into two segments on the $D6$ brane 
and the slow rotation of these two segments on the $67$ plane.
Later, the $D4$ brane starts to move towards the $D6$ brane faster and inflation ends. The final state configuration is described by the only supersymmetric
state given by $\Phi=0$ and $\phi_2=sqrt{\xi/g}$. This describes the $D4$ brane broken into two segments on the $D6$ brane and the two branes at the same $v$ coordinate which is the unique 
supersymmetric ground state. 

D--term inflation on D--branes: D--term inflation is very similar to the hybrid inflation. The only difference is the absence of the tree level inflaton mass which means that in the
initial configuration $\theta_1=0$ in addition to nonzero $\theta_2$ and $\theta_3$ as above. The scalar potential is
$$V=\lambda^2{(|\phi_1|^2+|\phi_2|^2)|\Phi|^2+\lambda^2 |\phi_1|^2|\phi_2|^2}+g^2(|\phi_1|^2-|\phi_2|^2+{\xi \over g})^2 \eqno(34)$$
The evolution is very similar to the previous case due to the inflaton mass generated at one--loop which arises from  
supersymmetry breaking during inflation
$$m^2= \xi^2 { g^2 \over {16 \pi^2}} {1 \over {\Phi^2}}={1 \over {16 \pi^2}} {\ell_s^5 \over { RL^4 v^2}} \eqno(35)$$
It was shown in ref. [\CAR] that the one--loop potential in eq. (10) is reproduced by the genus one string diagram in the above brane configuration.
This corresponds to a one loop potential between the $D4$ and $D6$ branes which is absent at tree level. This small, one loop potential leads
to the slow motion of the $D4$ brane towards the $D6$ brane.
The inflaton mass is now given by eq. (35) and inflation proceeds as in hybrid inflation. If $\theta_2=0$, then the Yukawa coupling is equal to 
the gauge coupling; $\lambda=g$. This corresponds to P--tem inflation[27] which is the ${\cal N}=2$ generalization of D--term inflation.
The parameters need to satisfy the slow--roll conditions (where $\lambda=g cos \theta_2$)
$$\epsilon_1={M_P^2 \over {2 \Phi^2}} log ({{\lambda \phi_2} \over M_P})={{2 \pi^2 M_P^2 \ell_s^4} \over v^2} log({{\lambda s} \over {2 \pi \ell_s^2 M_P}})<0.022 \eqno(36)$$ 
and
$$-0.06<\epsilon_2=-{M_P^2 \over \Phi^2} log ({{\lambda \phi_2} \over M_P})=-{{4 \pi^2 M_P^2 \ell_s^4} \over v^2} log({{\lambda s} \over {2 \pi \ell_s^2 M_P}})<0.05 \eqno(37)$$
We see that these require a large value for $\phi_2$ or large distances between the branes.
The magnitude of density perturbations is given by 
$$19 \times 10^{-10}<{1 \over {12 \pi^2}} {{\xi^2 \Phi^2} \over M_P^6} {1 \over {log ({{\lambda \Phi} \over M_P})}}={1 \over {3072 \pi^6}}{{L v^2 sin^2 \theta_3} 
\over {g_s \ell_s^9 M_P^6}} {1 \over { log({{\lambda v} \over {2 \pi \ell_s^2 M_P}})}} <25 \times 10^{-10} \eqno(38)$$
We also find
$$R={{32 \pi^2 M_P^2 \ell_s^4} \over v^2} log({{\lambda v} \over {2 \pi \ell_s^2 M_P}})<0.35 \eqno(39)$$
and $n_s=1$ up to $O(\epsilon_{1,2}^2)$. This last result is unique to D--term inflation on branes and implies a perfectly scale invariant spectrum for the scalar perturbations.

\bigskip

\centerline{\bf 5. Compactified D--Brane Constructions}

\medskip

As mentioned before, the main drawback of the D--brane inflation scenarios of the previous section is the fact that the models are not compactified down to four dimensions. In this
section we describe another brane construction which is compact and loosely related to the Hanany--Witten construction by T--duality. We generalize the results of ref. 
[\KAL] and describe the deformations of this new construction that correspond to the brane rotations in the Hanany--Witten model. 

First let us consider what happens to the Hanany--Witten construction under T--duality along a compact $6$ direction[\DAS] (which takes the string theory from IIA to IIB). The $D4$ brane 
becomes a $D3$ brane along $123$ directions
whereas the $D6$ brane becomes a $D7$ brane along $1236789$ directions. It is more difficult to see what happens to the two $NS5$ branes. T--duality turns the two $NS5$ branes into
five dimensional Kaluza--Klein monopoles of type IIB string theory. The geometry of two monopoles is given by the two center Taub--NUT space which is a good approximation to
($Z_2$) ALE space near the center of the ALE. Thus under T--duality the two $NS$ branes turn into pure geometry, i.e. ALE space along the $6789$ directions. Now it is easy to see how this 
configuration can be compactified along the $456789$ directions. The $45$ directions are transverse to both branes and can be compactified on a torus $T^2$. On the other hand,
the compactified version of ALE space is the $K3$ manifold. 

Consistency of the compactification requires the cancellation of global charges. In order to cancel the $D7$ charge we need to include orientifold planes which means that we need
to orientifold the $K3 \times T^2$. The correct compact manifold is given by $K3 \times T^2/Z_2$ where the orientifold operation is defined by $Z_2=\Omega (-1)^{F_L} 
{\cal I}_{45}$. Here ${\cal I}_{45}$ is the orbifold projection with the action $X_{4,5} \to -X_{45}$, $\Omega$ is the worldsheet orientation reversal and $(-1)^{F_L}$ changes the 
sign of the left moving fermions. There is one orientifold plane at each one of the four orbifold singularities of $T^2/{\cal I}_{45}$. 
The local charge is cancelled by placing
four $D7$ branes on top of each $O7$. Thus the compact model consists of 16 $D7$ branes and 4 $O7$ planes compactified on the orientifold $K3 \times T^2/Z_2$. We can
always separate one of the $D7$ branes from the other three at one of the fixed points and see how it behaves in the presence of a test $D3$ brane. The dynamics of this $D3$-$D7$
pair gives rise to inflation models under different deformations of this setup.

To summarize, the configuration with one $D3$ along $123$ directions and $4(O7+4D7)$ along $1236789$ directions on the compact space $K3 \times T^2/Z_2$ where the $D3-D7$ brane pair
is separated from the rest corresponds to the compactified version of the brane model based on the Hanany--Witten construction[\KAL]. This is a supersymmetric configuration and leads 
to inflation only under
one of the deformations which correspond to brane rotations of section 3. Therefore we need to find these deformations and show that they can be realized independently as in section
4.

Deformations of $T^2$:  This corresponds to the rotation of one of the $NS5$ branes relative to the other as in eq. (14). 
To understand this note that when the relative angle between the two
$NS5$ branes is $\pi/2$, the ${\cal N}=2$ SUSY of the field theory is broken to ${\cal N}=1$. This is accomplished from the field theory point of view by giving an infinite mass
to the singlet adjoint $\Phi$. We saw that this mass is $m=tan \theta_1/2 \pi \ell_s$. In ref. [\DAS] it has been shown that this ${\cal N}=1$ configuration corresponds to the $D3-D7$
brane system on a conifold which is described by the equation
$$(z_1)^2+(z_2)^2+(z_3)^3=-(z_4)^4 \eqno(35)$$  
where $z_4=X_4+iX_5$.
In this form the conifold is described as an ALE (along the $6789$ directions) space blown up by a sphere ($P^1$) of size $z_4$. In other words, the conifold can be considered as
a fibration where the base is the $z_4$ plane and the fiber is te ALE space. Thus, we find that the rotation of one of the $NS5$ branes corresponds the transformation of the torus 
$T^2$ into a sphere $P^1$. A small singlet mass which results in chaotic inflation would arise from a small deformation of the $T^2$. As we see it is very difficult to quantify 
this deformation and follow its effect on the scalar potential.

Mixing of the $K3$ and $T^2$: The common rotation of both $NS5$ branes is described by eq. (14) above which rotates the plane $45$ into the plane $89$.
Note that now the $NS5$ branes turned into pure geometry described by the space $K3 \times T^2/Z_2$ where the $K3$ and $T^2$ are along the $6789$ and $45$ directions respectively.
Thus, the common rotation of the $NS5$ branes corresponds to a mixing of the directions of the $K3$ and $T^2$. After the deformation $K3$ is along $67 w_{\theta}$ whereas $T^2$ is along
$v_{\theta}$. The Yukawa coupling vanishes when $K3$ and $T^2$ are along $4567$ and $89$ directions respectively. The maximun Yukawa coupling is $\lambda=g$ and is obtained when 
there is no rotation and $K3$ is along $6789$ and $T^2$ is along $45$ directions.

Flux on the $D7$ brane: The rotation of the $D6$ brane relative to the $D4$ brane in the Hanany--Witten configuration does not correspond to a deformation of the background geometry
since it does not involve the $NS5$ branes. In fact in ref. [\KAL] it was shown that it corresponds to a  nonself--dual gauge flux on the world--volume of the $D7$ brane. This is not
surprising if we remember that under T--duality a rotation tranforms into a world--volume gauge flux. Such a flux breaks SUSY and creates a force between the $D3$ and $D7$
branes. The relation between the flux and the anomalous D--term in the field theory is given by $g \xi=(\psi_1-\psi_2)/2 \pi$ where the constant world--volume gauge flux 
(${\cal F}=dA-B$) is
$${\cal F}_{67}=tan \psi_1 \qquad \qquad {\cal F}_{89}=tan \psi_2 \eqno(36)$$

Now these three deformation can be used to obtain the different models of inflation in a model compactified down to four dimensions. 
For example, if  $K3$ and $T^2$ are along $4567$ and $89$ directions respectively (so that the Yukawa coupling vanishes) and there is no flux on the $D7$ brane (so that there is no
anomalous D--term) but the $T^2$ is slightly deformed we get chaotic inflation. On the other hand, if the $T^2$ is not deformed and  $K3$ and $T^2$ are along $4567$ and $89$ directions 
respectively but there is a mixing among the directions of $K3$ and $T^2$ we get slow--roll inflation. If the $T^2$ is not deformed but the other two deformations are allowed we get
D--term inflation. Finally, if all deformations are allowed we get hybrid inflation.

In all cases the inflaton is the distance between the $D3$ and $D7$ branes. Inflation corresponds to the slow motion of the $D3$ brane towards the $D7$ brane which arises due to SUSY breaking
of the configurations. Note that this SUSY breaking may arise from the changes in the background geometry in addition to world--volume fluxes. The best argument for inflation
in these cases is our results in section 4 and T--duality. It is very difficult if not impossible to parametrize the deformations and to write down a scalar potential in terms of them.

\bigskip

\centerline{\bf 6. Conclusions and Discussion}

\medskip

In this paper, we obtained chaotic, slow--roll, hybrid and D--term inflation on D--branes by using the Hanany--Witten construction and its deformations. These deformations are given
by the relative angles between the different branes. As a result, all of the above models of inflation are continuously related to each other. The different inflation models are
obtained at different limits of the parameter space defined by the three angles $\theta_{1,2,3}$ in addition to the different values that define the string theory such as $g$,
$\ell_s$, $R$ and $L$. The crucial point is the fact that the deformations completely control the parameters that appear in the scalar potential, namely the inflaton mass $m$,
the Yukawa coupling $\lambda$ and the anomalous D--term $\xi$ allowing us to obtain different potentials at different limits.

The main drawback of the above construction based on the Hanany--Witten model is the fact that it is not compactified to four dimensions. However, as shown in ref. [13] and in section
5, there is another compactified construction which is loosely related to the previous one by T--duality and realizes the same models of inflation. Compactification comes at a price however. 
The deformations which are simple to visualize and easy to analyze in terms of a scalar potential turn into complicated geometrical deformations which cannot be easily analyzed.  
 
The Hanany--Witten construction and its deformations can be realized in M theory which leads to models of inflation on M branes. When lifted to 11 dimensions the $D4$ brane and the two
NS5 branes become one $M5$ brane on $R^4 \times Q$ where $Q$ is a two dimensional surface given by a Seiberg--Witten curve. The $D6$ brane when liften to 11 dimensions becomes a Taub--NUT 
space, the geometry in which the $M5$ brane 
is embedded. The deformations considered above, namely the rotations between the different branes correspond to algebraic deformations of the Seiberg--Witten curve. The inflaton which previously
corresponded to the relative distance or angle between the branes becomes deformation parameters of the Seiberg--Witten curve (or the
wrapped $M5$ brane). Inflation corresponds to the slow deformation of the shape of the NS5 brane and ends when the $M5$ brane wraps a supersymmtric cycle in $Q$ which gives a supersymmetric
configuration as the end point of inflation. It would be certainly worthwhile to see if the above construction can be realized by an $M5$ brane.

Any model of inflation on branes would also constitute a model for quintessence[\QUI,\STE] for a different choice of parameters. In this case, the parameters would have to be 
extremely fine--tuned due to the smallness of the present vacuum energy, e.g. $\xi \sim 10^{-60} M_P^2$.
It would be interesting to find out if any of the above models which lead to inflation can accomodate quintessence relatively naturally in a manner similar to the scenario in [\TEV].

Finally, the dS/CFT correspondence[\STR] allows us to analyze inflation in a holographic manner in terms of a CFT that flows to a fixed point[\INF, \HOL]. The CFT lives on the horizon of the
inflating space which is nearly de Sitter. Finding out whether inflation on D--branes sheds light on the dS/CFT correspondence and the nature of holography requires a detailed analysis.



\vfill

\refout

\end
\bye